\documentclass[12pt]{iopart}
\usepackage{iopams}  
\usepackage[dvips]{graphicx,color}
\usepackage{hyperref}  
\begin{document}

\title[Towards a controlled study of the QCD critical point]{
\hspace*{-0.2cm}Towards a controlled study of the QCD critical point
}

\author{Ph.~de Forcrand (in collaboration with O.~Philipsen)}

\address{
Institut f\"ur Theoretische Physik, ETH Z\"urich, CH-8093 Z\"urich, Switzerland, and \\
CERN, Physics Department, TH Unit, CH-1211 Geneva 23, Switzerland
%Institut f\"ur Theoretische Physik, Westf\"alische Wilhelms-Universit\"at M\"unster,
%D-48149 M\"unster, Germany
}
\ead{forcrand@phys.ethz.ch}
%ophil@uni-muenster.de
\begin{abstract}
The phase diagram of QCD, as a function of temperature $T$ and quark chemical
potential $\mu$, may  contain a critical point $(\mu_E,T_E)$ whose non-perturbative
nature makes it a natural object of lattice studies. However, the sign problem prevents
the application of standard Monte Carlo techniques at non-zero baryon density. We have 
been pursuing an approach free of the sign problem, where the chemical potential is taken
as imaginary and the results are Taylor-expanded in $\mu/T$ about $\mu=0$, then analytically
continued to real $\mu$.

Within this approach we have determined 
the sensitivity of the critical chemical potential $\mu_E$
to the quark mass, $d(\mu_E)^2/dm_q|_{\mu_E=0}$. Our study indicates that the critical point
moves to {\em smaller} chemical potential as the quark mass {\em increases}. This finding,
contrary to common wisdom, implies that the deconfinement crossover, which takes place in
QCD at $\mu=0$ when the temperature is raised, will remain a crossover in the $\mu$-region
where our Taylor expansion can be trusted. If this result, obtained on a coarse lattice,
is confirmed by simulations on finer lattices now in progress, then we predict that no
{\em chiral} critical point will be found for $\mu_B \lesssim 500$ MeV, unless the phase 
diagram contains additional transitions.
\end{abstract}

%\pacs{11.15.Ha, 12.38.Gc, 12.38.Mh, 25.75.Nq}

% \maketitle

\section{Motivation}
\vspace*{-15cm}
\begin{minipage}{\textwidth}
\begin{flushright}
\texttt{\footnotesize
CERN-PH-TH/2008-120
}
\end{flushright}
\end{minipage}\\[13.5cm]

QCD has a rich phase diagram as a function of quark chemical potential $\mu$ and tem-perature $T$,
with at least 3 regimes: confining (low $\mu,T$), quark-gluon plasma (high $T$) and color superconducting
(high $\mu$). Lattice simulations have provided clear evidence that, for $\mu\!\!=\!\!0$, the temperature-driven
"transition" between the first 2 regimes is actually a crossover~\cite{crossover}. Then, the commonly
expected $(\mu,T)$ phase diagram, depicted Fig.~1 (left), contains a critical point caused by the
$\mu\!=\!0$ crossover turning into a first-order transition. This critical point is the object of both
experimental search and theoretical lattice investigation. If located at small enough chemical 
potential, it could be discovered in high-energy, not-so-heavy ion collisions at RHIC or LHC. 
In fact, it has been predicted to lie at $(\mu_E,T_E) \!=\! (120(13),162(2))$ MeV in a celebrated lattice study~\cite{FK_crit}.

\begin{figure}[t]
\includegraphics[width=0.49\textwidth,height=5.5cm]{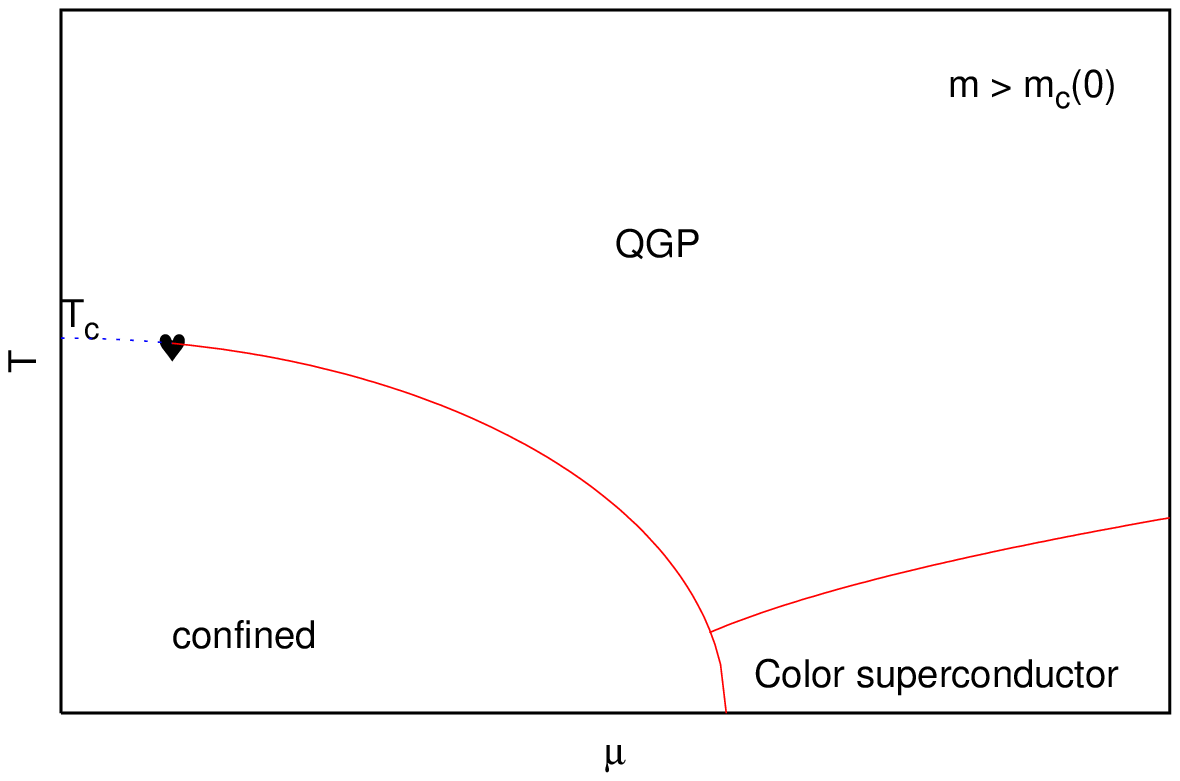}
\includegraphics[width=0.49\textwidth,height=5.1cm]{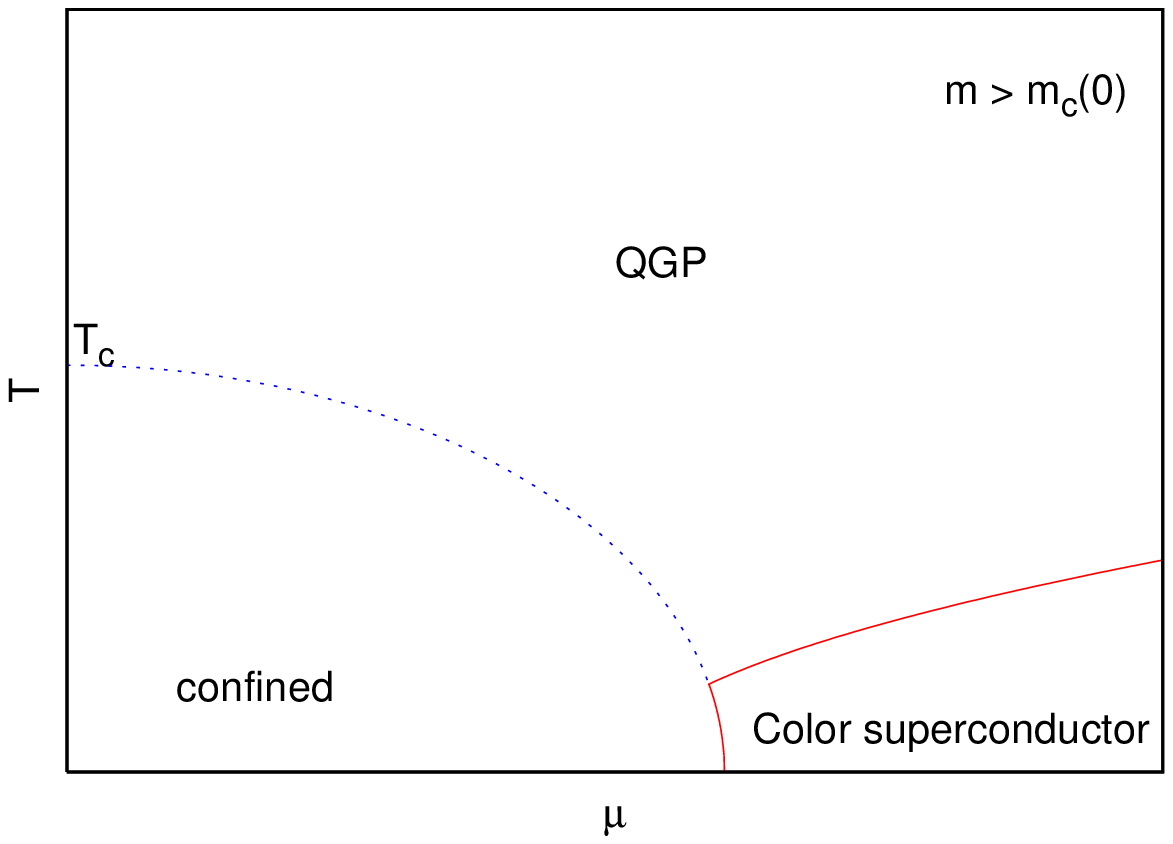}
\caption{Conventional (left) and simplest possible unconventional (right) phase diagram of QCD
as a function of chemical potential $\mu$ and temperature $T$. Dotted lines denote crossovers,
solid lines first-order phase transitions. The QCD critical point (end of first-order line),
marked by a $\heartsuit$ on the left, is absent on the right.
}
\label{fig1}
\end{figure}

However, the numerical method of Ref.~\cite{FK_crit}, which consists of reweighting results obtained from
$\mu\!=\!0$ simulations to $\mu\neq 0$, is known to fail and give wrong results without warning 
(see, e.g., \cite{Glasgow}) when the overlap between the $\mu\!=\!0$ Monte Carlo ensemble and the target $\mu\neq 0$
ensemble becomes insufficient. Therefore, the abrupt change observed in \cite{FK_crit} as a function of $\mu$
(see Fig.~1 there) might be caused by an abrupt breakdown of the numerical approach. In addition the needed
reweighting factors to the $\mu\neq 0$ target ensemble are not always positive, because of the notorious
``sign problem'' affecting the $\mu\neq 0$ determinant. This makes the statistical errors more difficult
to control. Finally, the lattices used in \cite{FK_crit} have $N_t\!=\!4$ time-slices, corresponding to a 
very coarse lattice spacing $a\sim 0.3$ fm. As acknowledged in \cite{FK_crit} already, the results may
change considerably after extrapolation to the continuum limit $a\to 0$. Therefore, the landmark
result of \cite{FK_crit} should not be considered the final word on the QCD critical point.
A more cautious crosscheck like the one we present here is warranted.

On the theoretical side, the expectation of a critical point as in Fig.~1 (left) can be traced back to
two assumptions: $(i)$ QCD with $N_f\!\!=\!\!2$ massless flavours, at $\mu\!\!=\!\!0$, has a second-order temperature-driven
phase transition in the $O(4)$ universality class~\cite{PW}. This implies, for this theory, the existence
of a tricritical point at $(\hat\mu,\hat{T})$, and for the $N_f\!\!=\!\!2+1$ theory that of another tricritical 
point at $(\mu\!\!=\!\!0,T^*,m_{u,d}\!=\!0,m_s^*)$. $(ii)$ These two tricritical points are analytically connected
by a line in the $(\mu,T,m_{u,d}\!=\!0,m_s)$ space.
However, assumption $(i)$ depends on the strength of the axial $U_A(1)$ symmetry breaking at $T_c$.
And assumption $(ii)$ has no other basis than simplicity.
This motivates us to take a careful look at the phase diagram of $N_f\!\!=\!\!2+1$ QCD, generalized to arbitrary quark masses
$m_u\!\! =\!\! m_d\!\! =\!\! m_{u,d}$ and $m_s$. We present our $\mu\!=\!0$, then $\mu\!\neq\! 0$ findings below.

\begin{figure}[t]
\includegraphics[width=0.43\textwidth]{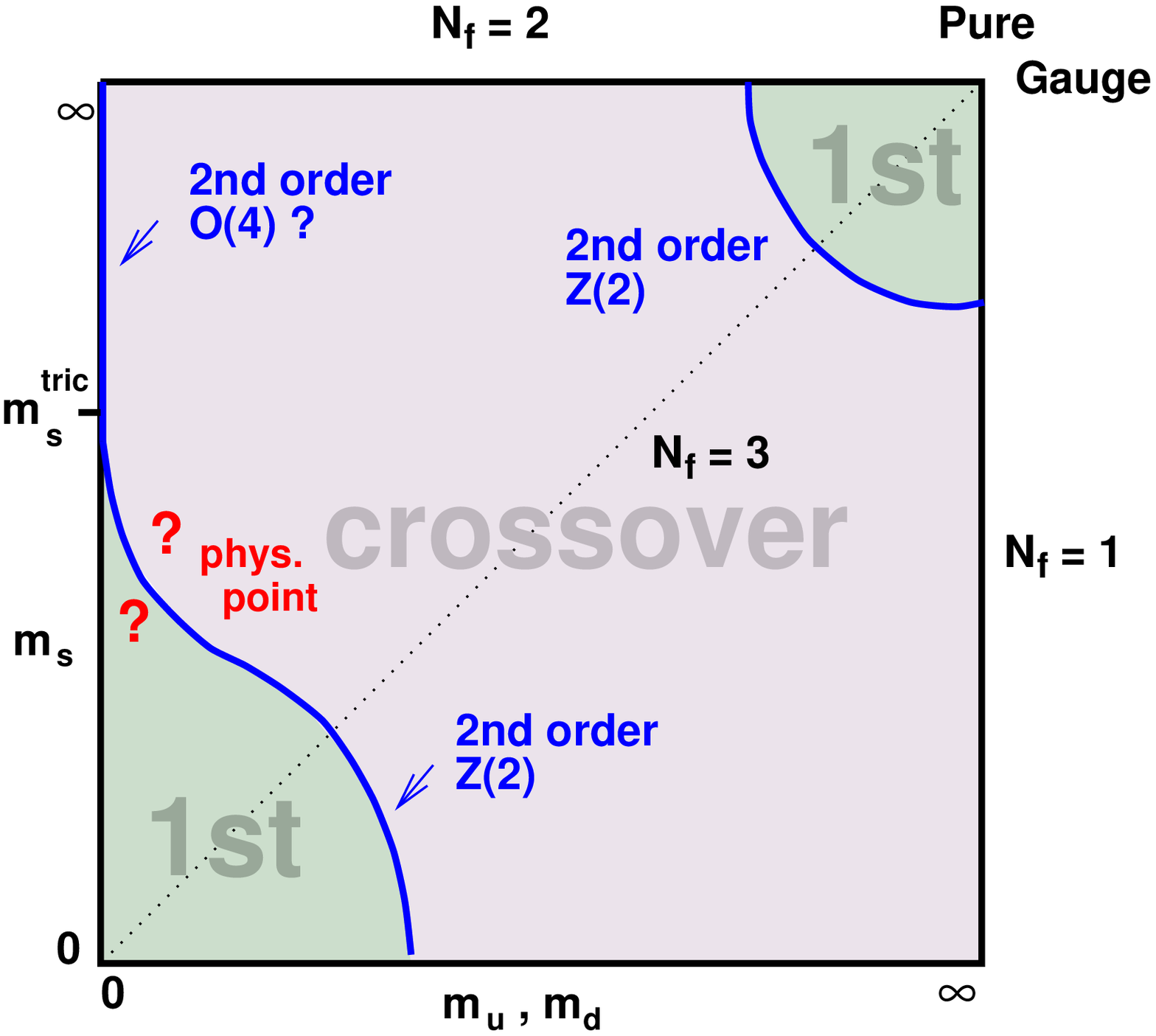}
\includegraphics[width=0.58\textwidth]{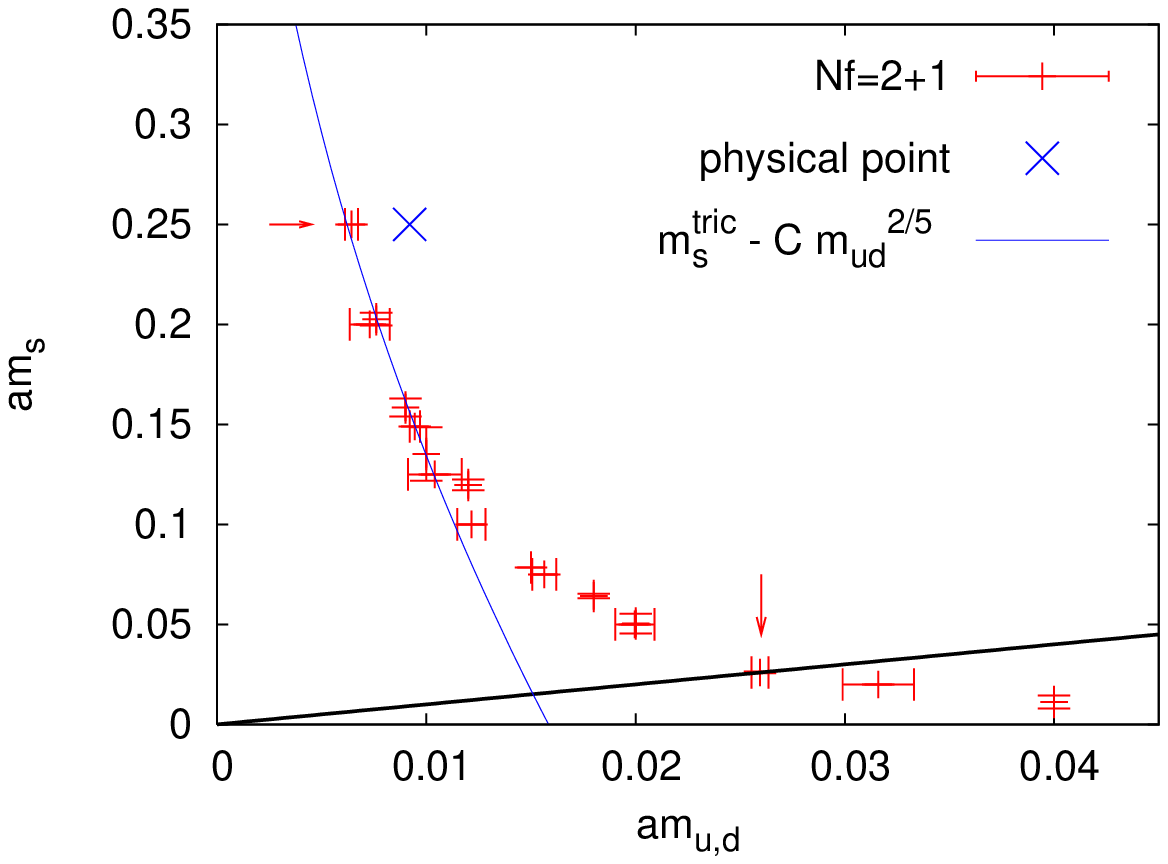}
\caption{Order of the temperature-driven phase transition (at $\mu\!=\!0$), as a function of the light
quark mass $m_{u,d}$ and the strange quark mass $m_s$. On the left, theoretical expectations~\cite{OP_review}.
On the right, $N_t\!=\!4$ lattice results for light quarks, from \cite{paper3}. The straight line corresponds
to $N_f\!=\! 3$ ($m_s\!=\!m_{u,d}$). The curve is the expected scaling behaviour from a tricritical point at $m_{u,d}\!=\!0,m_s\sim 500$ MeV.
The physical point, marked by a $\times$, is in the crossover region.
}
\label{fig2}
\end{figure}

\section{Approach}

To protect ourselves against the pitfalls of reweighting to $\mu\neq 0$, we consider the effect
of an {\em imaginary} chemical potential $\mu\!=\!i \mu_I$~\cite{paper12,paper3}.
Then the fermion determinant is positive
and no overlap problem occurs because we do not reweight.
The drawback is that, in order to analytically continue an observable, its 
$\mu$-dependence is 
fitted to low-order Taylor series about $\mu\!=\!0$, thus introducing a truncation error. 
As a safeguard, we also calculate the coefficients without truncations \cite{LAT07}.
Note that these derivatives can be expressed, and measured, 
as non-local observables in the $\mu\!=\!0$
ensemble. This is the strategy pursued successfully in \cite{B-S} for the pressure. Our approach
is computationally more economical, and we extend it to the critical line.

The first step is to determine, at $\mu\!=\!0$, the line in the $(m_{u,d}, m_s)$ plane for which 
the temperature-driven phase transition is second order. For lighter quarks, the transition becomes
first order as in the chiral limit; for heavier quarks, lattice simulations show a crossover.
Theoretical expectations are displayed in Fig.~2 (left). Our determination of the critical line
on an $N_t\!=\!4$ ($a\sim 0.3$ fm) lattice is shown Fig.~2 (right), and matches expectations, 
including the possible scaling behaviour in the vicinity of a tricritical point at 
($m_{u,d}\!=\!0,m_s^{\rm tric}\sim 500$ MeV). Also, the physical point (with real-world $\pi$ and $K$ meson masses)
lies, as expected, in the crossover region.

We now choose a point on the critical line just determined, and monitor the effect of a small
$\mu$. This exercise has been performed, with similar results, at the two points marked
by arrows in Fig.~2 (right), with higher accuracy for the $N_f\!\!=\!\!3$ case. If conventional wisdom is right,
then the surface spanned by the critical line as $\mu$ is turned on will bend towards the crossover
region, so as to produce a critical point at some small $\mu$ value for physical quark masses.
This is depicted Fig.~3 (left). On the contrary, an opposite curvature will indicate the absence of
a critical point at small $\mu$, where our truncated Taylor expansion is a good approximation. 
This is illustrated Fig.~3 (right).
The issue is thus to determine whether the $\mu\!=\!0$ second order transition becomes first order or
crossover for a small $\mu$. We have now confirmed our first study~\cite{paper3} by a second one
using higher statistics and an improved numerical method~\cite{LAT07}. We measure the change in the Binder cumulant
$B_4(X) \equiv \langle (X - \langle X \rangle)^4 \rangle / \langle (X - \langle X \rangle)^2 \rangle^2$,
with $X \!=\! \bar\psi \psi$. At the second order transition, $B_4$ takes the value 1.604 dictated by the 
$3d$ Ising universality class. Reweighting with a small imaginary $\mu$ induces small changes in $B_4$,
which can be measured accurately because statistical errors largely cancel between the reweighted and
original ensembles.
Our current result for the $N_f\!\!=\!\!3$ ($m_{u,d}\!=\!m_s$) theory is
\begin{equation}
\frac{m_c(\mu_q)}{m_c(0)} = 1 {\bf -} 3.3(5) \left(\frac{\mu_q}{\pi T}\right)^2
{\bf -} 12(6) \left(\frac{\mu_q}{\pi T}\right)^4
+ \dots
\end{equation}
supporting the unconventional Fig.~3 (right), both in the leading and 
subleading term.

\begin{figure}
\hspace*{-0.7cm}
\includegraphics[width=0.55\textwidth]{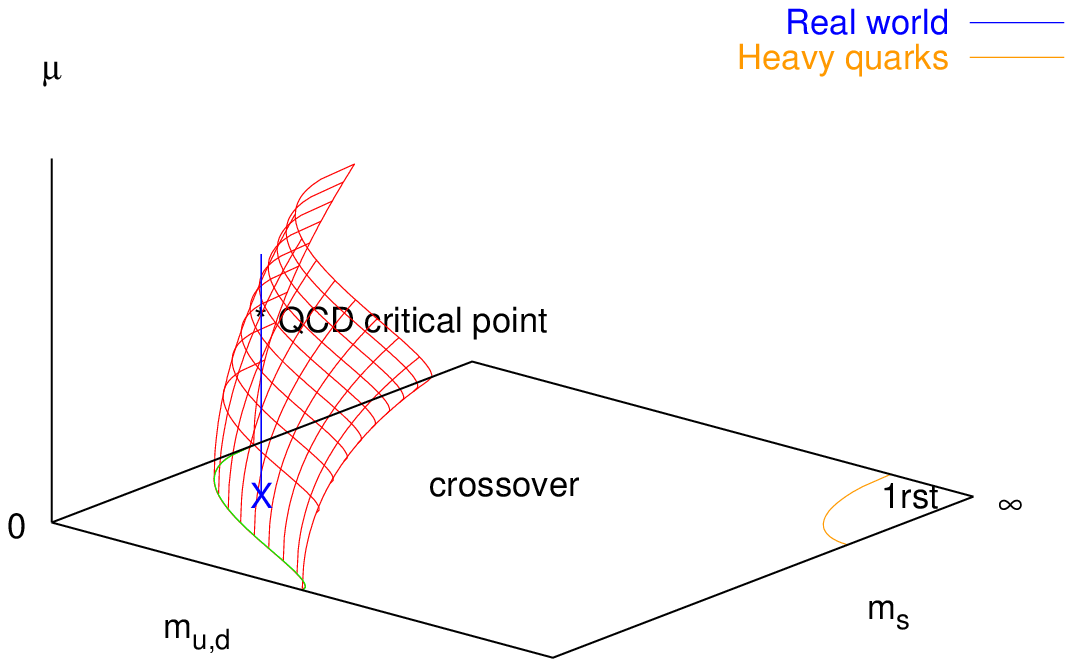}
\put(-174.9,88){$\bullet$}
\put(-182,88){\color{blue}\bf $\uparrow$}
\put(-180,49){\color{red} $\bullet$}
\put(-195,49){\color{blue}\bf $\longleftarrow$}
\includegraphics[width=0.55\textwidth]{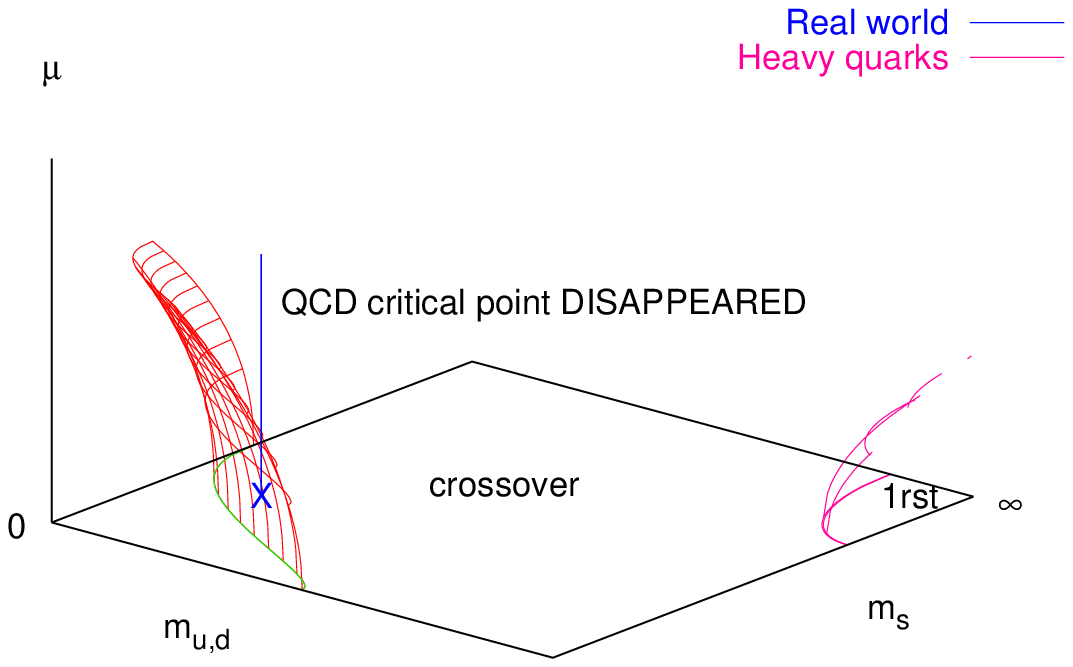}
\put(-180.5,49){\color{red} $\bullet$}
\put(-195,49){\color{blue}\bf $\longleftarrow$}
\caption{Critical surface spanned by the critical line of Fig.~2 as $\mu$ is turned on.
Depending on the curvature of this surface, a QCD critical point is present (left) or
absent (right). For heavy quarks, the curvature (right) has been determined in \cite{Potts}.
The arrows indicate the effect of a finer lattice: the distance between the physical quark
masses and the critical surface increases, driving the critical point (left) to larger $\mu$.
}
\label{fig3}
\end{figure}

It is now mandatory to repeat this study on a finer lattice, in order to gain some control
over the continuum limit. $N_t\!=\!6$ ($a\sim 0.2$ fm) simulations are in progress.
The first step, at $\mu\!=\!0$, already reveals an important shift of the critical line
towards the origin: for the $N_f\!=\!3$ theory, the pion mass,
measured at $T\!=\!0$ with the critical quark mass, decreases from $1.6\; T_c$ to $0.95\; T_c$~\cite{LAT07}.
As shown Fig.~3, this considerably increases the distance of the critical surface to the
physical point, pushing the critical point to larger $\mu$ (left), or requiring larger higher-order
terms of the right sign to bend the critical surface ``back'' (right). Regardless of the sign of the 
curvature in the continuum limit, this trend alone makes a QCD chiral critical point at small
$\mu_q/T \lesssim 1$ unlikely.

\end{document}